\documentclass[twocolumn,showpacs,preprintnumbers,amsmath,amssymb,aps,prc,floatfix,letterpaper,superscriptaddress]{revtex4}

\usepackage{graphicx}
\usepackage{dcolumn}
\usepackage{bm}

\newcommand\NSU{Norfolk State University, Norfolk, Virginia 23504}
\newcommand\UVa{University of Virginia, Charlottesville, Virginia 22903}
\newcommand\Jlab{Thomas Jefferson National Accelerator Facility,
  Newport News, Virginia 23606 }
\newcommand\UBasel{Universit\"{a}t Basel, CH-4056 Basel, Switzerland}
\newcommand\FIU{Florida International University, Miami, Florida 33199}
\newcommand\HamptonU{Hampton University, Hampton, Virginia 23668}
\newcommand\MissSU{Mississippi State University, Mississippi State,
  Mississippi 39762}
\newcommand\NCAT{North Carolina A\&T State University, Greensboro,
  North Carolina 27411}
\newcommand\ODU{Old Dominion University, Norfolk, Virginia 23529}
\newcommand\SUNO{Southern University at New Orleans, New Orleans,
  Louisiana 70126}
\newcommand\TelAviv{Tel Aviv University, Tel Aviv, 69978 Israel}
\newcommand\UMD{University of Maryland, College Park, Maryland 20742}
\newcommand\UNCW{University of North Carolina, Wilmington, North Carolina 28403}
\newcommand\VaTech{Virginia Polytechnic Institute \& State University,
Blacksburg, Virginia 24061}
\newcommand\Yerevan{Yerevan Physics Institute, Yerevan, Armenia}

\begin{document}


\title{ Proton $G_E$/$G_M$ 
from beam-target asymmetry }


%
\author{M.~K.~Jones}        \affiliation{\Jlab}
\author{A.~Aghalaryan}      \affiliation{\Yerevan}
\author{A.~Ahmidouch}        \affiliation{\NCAT}
\author{R.~Asaturyan}        \affiliation{\Yerevan}
\author{F.~Bloch}        \affiliation{\UBasel}
\author{W.~Boeglin}        \affiliation{\FIU}
\author{P.~Bosted}        \affiliation{\Jlab}
\author{C.~Carasco}        \affiliation{\UBasel}
\author{R.~Carlini}        \affiliation{\Jlab}
\author{J.~Cha}         \affiliation{\MissSU}
\author{J.P.~Chen}        \affiliation{\Jlab}
\author{M.E.~Christy}        \affiliation{\HamptonU}
\author{L.~Cole}        \affiliation{\HamptonU}
\author{L.~Coman}        \affiliation{\FIU}
\author{D.~Crabb}        \affiliation{\UVa}
\author{S.~Danagoulian}     \affiliation{\NCAT}
\author{D.~Day}         \affiliation{\UVa}
\author{J.~Dunne}        \affiliation{\MissSU}
\author{M.~Elaasar}        \affiliation{\SUNO}
\author{R.~Ent}         \affiliation{\Jlab}
\author{H.~Fenker}        \affiliation{\Jlab}
\author{E.~Frlez}        \affiliation{\UVa}
\author{D.~Gaskell}                \affiliation{\Jlab}
\author{L.~Gan}                \affiliation{\UNCW}
\author{J.~Gomez}        \affiliation{\Jlab}
\author{B.~Hu}          \affiliation{\HamptonU}
\author{J.~Jourdan}        \affiliation{\UBasel}
\author{C.~Keith}        \affiliation{\Jlab}
\author{C.E.~Keppel}        \affiliation{\HamptonU}
\author{M.~Khandaker}        \affiliation{\NSU}
\author{A.~Klein}        \affiliation{\ODU}
\author{L.~Kramer}        \affiliation{\FIU}
\author{Y.~Liang}        \affiliation{\HamptonU}
\author{J.~Lichtenstadt}    \affiliation{\TelAviv}
\author{R.~Lindgren}        \affiliation{\UVa}
\author{D.~Mack}        \affiliation{\Jlab}
\author{P.~McKee}        \affiliation{\UVa}
\author{D.~McNulty}        \affiliation{\UVa}
\author{D.~Meekins}        \affiliation{\Jlab}
\author{H.~Mkrtchyan}        \affiliation{\Yerevan}
\author{R.~Nasseripour}     \affiliation{\FIU}
\author{I.~Niculescu}        \affiliation{\Jlab}
\author{K.~Normand}        \affiliation{\UBasel}
\author{B.~Norum}        \affiliation{\UVa}
\author{D.~Pocanic}        \affiliation{\UVa}
\author{Y.~Prok}        \affiliation{\UVa}
\author{B.~Raue}        \affiliation{\FIU}
\author{J.~Reinhold}        \affiliation{\FIU}
\author{J.~Roche}        \affiliation{\Jlab}
\author{D.~Rohe}        \affiliation{\UBasel}
\author{O.A.~Rond\'{o}n}        \affiliation{\UVa}
\author{N.~Savvinov}        \affiliation{\UMD}
\author{B.~Sawatzky}        \affiliation{\UVa}
\author{M.~Seely}        \affiliation{\Jlab}
\author{I.~Sick}        \affiliation{\UBasel}
\author{K.~Slifer}         \affiliation{\UVa}
\author{C.~Smith}        \affiliation{\UVa}
\author{G.~Smith}        \affiliation{\Jlab}
\author{S.~Stepanyan}        \affiliation{\Yerevan}
\author{L.~Tang}        \affiliation{\HamptonU}
\author{S.~Tajima}         \affiliation{\UVa}
\author{G.~Testa}        \affiliation{\UBasel}
\author{W.~Vulcan}        \affiliation{\Jlab}
\author{K.~Wang}        \affiliation{\UVa}
\author{G.~Warren}        \affiliation{\UBasel}\affiliation{\Jlab}
\author{F.R.~Wesselmann}    \affiliation{\UVa}\affiliation{\NSU}
\author{S.~Wood}        \affiliation{\Jlab}
\author{C.~Yan}         \affiliation{\Jlab}
\author{L.~Yuan}        \affiliation{\HamptonU}
\author{J.~Yun}         \affiliation{\VaTech}
\author{M.~Zeier}        \affiliation{\UVa}
\author{H.~Zhu}         \affiliation{\UVa}

\collaboration{The Resonance Spin Structure Collaboration}
\noaffiliation

\date{\today}

\begin{abstract}
The ratio of the proton's electric to magnetic form factor, $G_E/G_M$, can
be extracted in elastic electron-proton scattering 
by measuring either cross sections, beam-target asymmetry or
recoil polarization.
Separate determinations of $G_E/G_M$ by cross sections and 
recoil polarization observables disagree for $Q^2 > 1$~(GeV/c)$^2$.
Measurement by a third technique might uncover an unknown systematic error
in either of the previous measurements.
The beam-target asymmetry has been measured for
elastic electron-proton scattering at $Q^2$ = 1.51~(GeV/c)$^2$ for
target spin orientation aligned perpendicular to the
beam momentum direction. This is the largest $Q^2$ at which 
$G_E/G_M$ has been determined by a  beam-target asymmetry experiment.
The result, $\mu G_E/G_M$ $= 0.884 \pm 0.027 \pm 0.029$, is compared to
previous world data.

\end{abstract}

\pacs{25.30.Bf,13.40.Gp}
\maketitle
\section{Introduction}

Understanding the structure of the nucleon has long been a goal of
nuclear physics and elastic electron-nucleon scattering has been an important
tool in this quest. In the one-photon exchange (Born) approximation,
the structure of the nucleon  can be characterized in terms of 
the electric and magnetic  form factors, $G_E$ and $G_M$, which depend only
on the four-momentum transfer squared, $Q^2 = -t$. At $Q^2 = 0$, 
the proton form factors are defined as  $G_E = 1$ and
$G_M = \mu$, where $\mu = 2.7928$ is the proton's magnetic moment.
The proton form factors can be 
extracted individually in elastic electron-proton scattering by measuring cross sections 
at the same $Q^2$ but different beam energies (Rosenbluth technique). 
In addition, spin observables in elastic electron-proton
scattering are sensitive to the ratio of $G_E$ to $G_M$. 

Historically, the Rosenbluth technique was used to measure $G_E$ and $G_M$
with elastic scattering identified by detection of the scattered electron.
The cross section can be written as:
\begin{eqnarray}
\frac{d\sigma}{d\Omega} &=& \frac{\alpha^2 E^{\prime} \cos^2\frac{\theta_e}{2}}{4(1+\tau)E^{3}\sin^4\frac{\theta_e}{2}}
\left[G^{2}_{E} + \frac{\tau}{\epsilon}G^{2}_{M}\right] \label{eq:xn} \\
\tau &=& \frac{Q^2}{4M^2} \,\,\,\,\,\,\,\, Q^2 = 2EE^{\prime}(1-\cos\theta_e)  \nonumber\\
\epsilon &=& \left[1+2(1+\tau)\tan^2\frac{\theta_e}{2}\right]^{-1} \nonumber
\end{eqnarray} 
where $E$ and $E^{\prime}$ are the incoming and outgoing electron energies, $M$ is the proton mass
and $\theta_e$ is the outgoing electron's scattering angle.
$G_M$ in Eq.~\ref{eq:xn} is multiplied by $Q^2$ and  dominates the cross sections at large $Q^2$
at all $\epsilon$. 
For example, at $Q^2 = 6$~(GeV/c)$^2$, the contribution of $G_E$ to the elastic cross section
is 7\% at $\epsilon = 0.9$, assuming $\mu G_E/G_M = 1$. 
 
\begin{figure}[tbh]
\includegraphics[width=\columnwidth]{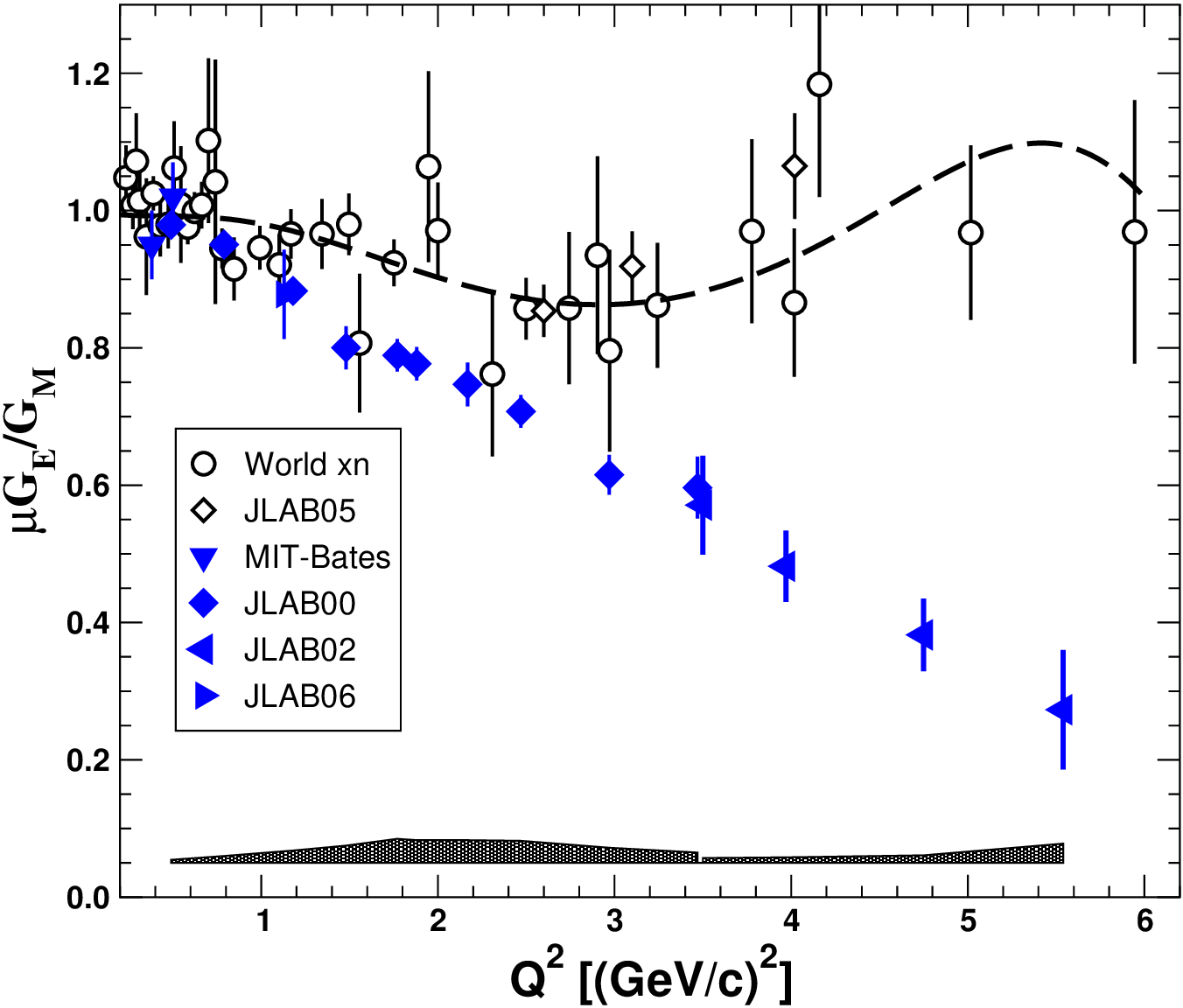}
\caption{
\label{fig:rat1} (Color online)
Ratio $\mu G_E/G_M$  plotted as a function of $Q^2$. 
``World xn'' and JLAB05~\cite{Qa05} used the Rosenbluth technique. 
Results using the recoil polarization technique are from MIT-Bates~\cite{Mi98}, JLAB00~\cite{Jo00},
JLAB02~\cite{Ga02} and JLAB06~\cite{Ma06}. The band at the bottom is the systematic error
on the data from JLAB00 and JLAB02. The dashed curve is a recent fit~\cite{Ar04d} to the world
cross section data.
}
\end{figure}

At SLAC, $G_E/G_M$ was measured to 
$Q^2 = 8.8$~(GeV/c)$^2$ using the Rosenbluth technique \cite{An94}. 
A recent JLab Hall~C experiment~\cite{Ch04} in the same $Q^2$ range agrees with
the SLAC data. These data were combined together with other cross sections
measurements for a global analysis by Arrington~\cite{Ar04d}. 
The $\mu G_E/G_M$ extracted from the global analysis is 
plotted in Fig.~\ref{fig:rat1} and labeled
``World xn''. The dashed line in 
Fig.~\ref{fig:rat1} is $\mu G_E/G_M$ from a fit by Arrington
to  that data with a polynomial parametrization of $G_E$ and $G_M$. 

Previous cross sections measurements detected electrons
to identify an elastic event.
A recent JLab experiment~\cite{Qa05} in Hall~A identified an elastic
scattering event by detection of the scattered proton. This experimental approach
has different systematic errors compared to electron detection and
has many advantages in terms of reducing the systematic error. 
The $\mu G_E/G_M$ are plotted 
in  Fig.~\ref{fig:rat1} and labeled as ``JLab05''. The new data agree well with the
recent fit to previous world data which demonstrates that the systematic errors
in the Rosenbluth technique are understood.

Early on, it was proposed~\cite{Ak68,Do69,Ak74} that measuring polarization observables in 
elastic electron-proton scattering would be an alternative 
method to extract the electric form factor given that the dominant magnetic
form factor is determined by cross section data. In 1976, an experiment~\cite{Al76}
measured the beam-target asymmetry for elastic $ep$ scattering 
at $Q^2 = 0.76$~(GeV/c)$^2$. But given that the experiment used a
longitudinally polarized target, the asymmetry was extremely insensitive to
$G_E/G_M$ and could only restrict the relative sign between $G_E$ and $G_M$.

With the advent of high duty factor, high current, and highly polarized
electron beam accelerators such as Jefferson Lab and the Mainz Microtron,
experiments which measure the proton and neutron electro-magnetic form factors
have reached a new level of precision over a larger $Q^2$ range by measuring
polarization observables in elastic electron-nucleon scattering (see Ref.~\cite{Hy04} for
a review of the recent experiments). The proton $G_E/G_M$ ratios have been extracted from
measurement of the recoil polarization components of the scattered protons in
elastic scattering of polarized electrons from an unpolarized proton target.
Both the transverse, $P_x$, and longitudinal, $P_z$, components of scattered proton's
recoil polarization are dependent on $G_E/G_M$. By simultaneously measuring
both components, one can extract $G_E/G_M$ from the ratio of polarization
components, $P_x/P_z$,  which cancels systematic errors from the beam polarization and
the analyzing power. 

The first measurements of $G_E/G_M$ using the polarization
transfer technique were done at MIT-Bates~\cite{Mi98} in the 1990's at $Q^2 = 0.38$ and 0.5~(GeV/c)$^2$
and are plotted in Fig.~\ref{fig:rat1}. The results agree
with $G_E/G_M$ from the Rosenbluth technique. 
The polarization transfer technique was used in Hall A at Jefferson
Lab~\cite{Jo00,Ga02} to measure $G_E/G_M$ to $Q^2 =$5.6~(GeV/c)$^2$ and the data 
are plotted in Fig.~\ref{fig:rat1}.   
A linear fall-off with $Q^2$ is seen which  is in sharp contrast to the nearly flat $Q^2$ dependence 
of $G_E/G_M$ measured with the Rosenbluth technique. The absolute systematic error on the polarization
transfer technique is given by the solid band at the bottom of Fig.~\ref{fig:rat1}. Reconciling 
the $G_E/G_M$ results from the two techniques is impossible given
the systematic error quoted for both techniques. 
A recent result~\cite{Ma06} using the polarization transfer technique in Hall~C at Jefferson Lab
for $G_E/G_M$   at  $Q^2 =$1.13~(GeV/c)$^2$  is plotted 
in  Fig.~\ref{fig:rat1} with the error bar
that is dominated by statistics. 

One possible solution that reconciles the different $G_E/G_M$ from the two experimental 
techniques is inclusion of two-photon exchange mechanisms which are not part of the standard radiative
correction procedure which reduces the raw cross section data to the 
Born cross sections needed in Eq.~\ref{eq:xn} 
for determination of $G_E$ and $G_M$.  The Coulomb distortion effect is one type of two-photon exchange mechanism 
(exchange of  one hard and one soft photon) which has been neglected in $ep$ experiments.
Calculations~\cite{Ar04c} which include Coulomb distortion effects when
 extracting the form factors from the cross sections find that $\mu G_E/G_M$ is 
reduced by about 0.05 for $Q^2 > 1$~(GeV/c)$^2$ while the effect on $\mu G_E/G_M$ is gradually reduced
at smaller $Q^2$. 

More general calculations~\cite{Bl05,Af05} of the contribution of two-photon exchange 
mechanisms in elastic electron-proton scattering have been done. The two calculations take 
different approaches to the model of the nucleon
which is needed as part of the two-photon exchange calculation. The approach of 
Ref.~\cite{Bl05} is applicable to lower $Q^2$ then that of Ref.~\cite{Af05}.
In both calculations, the contribution of the two-photon exchange amplitude has an $\epsilon$-dependence which 
has the same sign as the $G_E$ contribution to the cross section 
and is large enough to effect the extracted value of  $G_E$.
Therefore, the extracted  $G_E/G_M$ for the Rosenbluth technique is
reduced. 

In addition to a linear $\epsilon$-dependence,  both calculations have a 
nonlinear $\epsilon$-dependence in the two-photon
contribution to the cross section. A global analysis~\cite{Tvaskis:2005ex} of 
the $\epsilon$-dependence of elastic and inelastic cross sections found 
that the elastic ( inelastic) data was consistent with a maximum 
deviation from a linear fit of $\leq 0.4\%$ ($0.7\%$). But this level of precision 
is obtained by averaging over $0.2 < Q^2 < 5.2$~(GeV/c)$^2$ range. Since the amount of
nonlinearity can change with $Q^2$, more precise data is needed for comparison to theory.    
An approved JLab experiment~\cite{Ar04} is an extensive study of  non-linearity 
in the $\epsilon$-dependence of the elastic electron-proton 
cross section at fixed $Q^2$ for a number of different $Q^2$.

The effect of  two-photon exchange amplitude on the
polarization components is small, though the size of the contribution changes with $\epsilon$.
The recoil polarization measurements at JLab ran at $\epsilon$ between 0.45 and 0.77. 
From Ref.~\cite{Bl05}, the measured $P_x/P_z$ at $\epsilon = 0.5$ 
should be reduced by factors of  0.9975 and 0.97 
at $Q^2 = 1$ and 6~(GeV/c)$^2$, respectively. At $Q^2 = 3$~(GeV/c)$^2$, Ref.~\cite{Bl05} predicts that 
$P_x/P_z$  will be 4\% larger at $\epsilon = 0.05$ compared to $\epsilon = 1$  due to contributions from
two-photon amplitudes. Interestingly, the calculation of Ref.~\cite{Af05} predicts that  $\epsilon$-dependence of
 $P_x/P_z$ will have a slope of the opposite sign. 
Complementing the approved JLab cross section
experiment, an upcoming
JLab experiment~\cite{Gi04} will measure the $\epsilon$-dependence of $P_x/P_z$ at $Q^2= 2.6$~(GeV/c)$^2$. 

The two-photon models need to be tested by comparing  predictions of
additional observables to data. Experiments planned at Jefferson Lab~\cite{Af04} and proposed 
at VEPP-3 in Novosibirsk~\cite{Ar04b} would precisely measure 
the $\epsilon$-dependence of the ratio of cross sections, $R_{e+e-}$, 
for elastic electron-proton scattering to positron-proton
scattering at a fixed $Q^2$.  In absence of two-photon
mechanisms, the ratio would be one and independent of beam energy.
The present data set for $R_{e+e-}$ is limited with most measurements at $\epsilon > 0.6$.
Previous experimental data was re-examined~\cite{Arrington:2003ck} and found that
combining all data for $Q^2 < 2$~(GeV/c)$^2$ gives  a slope of -5.7 $\pm$ 1.8 \% for
the  $\epsilon$-dependence of $R_{e+e-}$. Indeed,
the calculation of Ref.~\cite{Bl05} predicts an $\epsilon$-dependence which is
consistent with the large error bars of the existing data.

Checking on the possibility of an unknown systematic error in the Rosenbluth or recoil polarization technique
is also important.  Measurement of  the beam-target asymmetry in elastic electron-proton scattering
offers an independent technique of determining $G_E/G_M$. The systematic errors are different
when compared to either the Rosenbluth technique or the polarization transfer technique. 
For elastic scattering, the recoil polarization of scattered proton is directly related to
the beam-target asymmetry by time reversal invariance. Therefore,
sensitivity of the beam-target asymmetry to two-photon effects is the same as in the 
recoil polarization technique. By measuring $G_E/G_M$ by a third technique 
and comparing to previous results, the discovery
of unknown or underestimated systematic errors in the previous measurements is possible. 

\section{Experimental Set-up}

The experiment was performed in Hall C at the Thomas Jefferson 
National Accelerator Facility (Jefferson Lab).
The main purpose of the experiment was a measurement of the inclusive
parallel and perpendicular spin asymmetries in the resonance region
for electron scattering on  polarized proton and deuterium targets. This report 
presents a subset of the data which measured the  perpendicular 
beam-target asymmetry for elastic electron-proton scattering.

Polarized electrons with 5.755~GeV/c momentum were scattered 
from polarized frozen ammonia ($^{15}$NH$_3$) with the spin of the 
polarized target aligned perpendicular to the beam.
The scattered electrons were detected at 13.15$^{\circ}$ 
in the High Momentum Spectrometer (HMS) which was set at a
central momentum of 4.73~GeV/c.  Electron particle identification was done
by a combination of a gas Cerenkov detector and lead-glass calorimeter.
A cut was placed to use a momentum range of  $\pm$8\%.

The frozen ammonia target~\cite{Cr95}  is  polarized  by dynamic nuclear polarization and 
operated at 1~K in a 5~T magnetic field. The magnetic field is created by a pair of  
superconducting Helmholtz coils which produces a uniform magnetic field that selects
the spin direction of the protons. The refrigerator is a $^4$He evaporation type which
is installed vertically along the center of the magnet.
The coils can be rotated independently of the refrigerator so 
that the target spin can be aligned to any angle relative to the beam.
The angle of the coils relative to the beam was measured to a precision of 0.1$^{\circ}$.

To make the target, frozen ammonia is pulverized 
into small fragments which are sifted to get fragments of the same size. 
The fragments are stored in sample bottles in liquid nitrogen dewars. For use in the
experiment, the ammonia fragments are placed in a cylindrical container which is
3~cm long with a diameter of 2.5~cm. Inside the container is a coil for measuring the
NMR signal.  The container is placed on an insert ladder so that the beam passes through
the container lengthwise. The insert ladder can be rotated independently of the magnet coils
and refrigerator so that the beam enters the container perpendicular to its face. 
To check the orientation of the insert ladder, a target was placed on the insert  which
consisted of L-shaped rods of tungsten separated by 3cm. From reconstruction of
the rods, the insert ladder was determined to be rotated 6$^{\circ}$ relative to the beam direction. 
The insert ladder held two frozen ammonia containers which were designated as TOP
and BOTTOM. Additional targets on the insert were a 6.9-mm-thick $^{12}$C disk and an empty container.
The targets are in a bath of liquid helium that is cooled by the refrigerator.

To maintain reasonable target polarization, the beam current was limited to 100~nA and
was uniformly rastered. The uniformity of the raster was obtained by
independently and simultaneously rastering at a fast frequency (17.9~kHz in vertical
direction and 24.2~kHz in the horizontal direction) over 1~mm square spot 
and slow frequency (30~Hz) over 0.9~cm maximum radius spiral pattern. The
slow raster frequency was the same frequency as the flipping of the beam helicity.
Each of the rasters could 
independently be turned on or off and the raster size changed. 
The beam position was measured on an
event-by-event basis using an array of secondary emission monitors~\cite{St00} located
upstream of the target.

At thermal equilibrium at 5~T and 1~K, the protons have a small polarization of 0.51\% 
and the electrons have a large polarization of 99.8\%.
By applying a microwave radiation to the target material
at a frequency near the electron spin-flip resonance frequency, the electron polarization
is transferred to the proton. The protons have a slow relaxation time compared to the electrons
and slowly the polarization of the protons builds up.
The spin vector of the  polarized protons is aligned parallel or anti-parallel to the field direction
by changing the frequency of the microwaves and measurements were done at both microwave
frequencies. For this data set, the target field
was aligned at 90$^{\circ}$ to the beam direction with positive target polarization defined
as the target field pointing toward beam left.
The target polarization slowly decreased with exposure to the beam. When it became too
small, the target was retracted from the beam to be annealed and repolarized.

The target polarization, $P_T$, was measured  by the NMR technique.  
To extract absolute polarization, the NMR signal was calibrated by a known
polarization at thermal equilibrium with no microwave
radiation and no beam. Under these conditions, the proton polarization can be accurately calculated
and used to determine the calibration constant, $C_{TE}$, of the NMR signal. 
$C_{TE}$ was determined separately for the bottom and top target,
since each target has an individual NMR setup. The normalization was taken from the weighted
average of a series of thermal equilibrium (TE) measurements which gives a small statistical error on $C_{TE}$. 
To determine the systematic error on $C_{TE}$, three separate series of TE measurements were done for
one target at different times and the standard deviation was found to be 2.9\%. This was used as 
the relative systematic error on the target polarization for both targets.

The accelerator at Jefferson Lab 
produces highly polarized beam that can be simultaneously delivered to all three experimental halls. 
The polarized beam was produced by photo-emission from a semiconductor cathode using
polarized laser light from a pulsed diode laser. Each hall had its own diode laser which produces
a narrow pulse, but a small continuous noise was also present. This produced a leakage current
from the other hall's laser underneath the main beam pulse for that hall. The leakage current 
was measured in Hall~C by an intrusive method. The rate in the HMS was measured with the Hall~C
laser turned on (normal conditions) and turned off (only leakage current). 
The ratio of the two rates is a measure of the leakage current. Throughout the experiment 
the leakage current was measured every 12 hours and
on average the leakage current was found to be 2\% of the total current. This is
the leakage current from both Halls~A and B.

The polarization of electrons produced at the cathode depends on the laser wavelength. 
At the time of this experiment, Hall~A wanted high current and was not 
interested in polarized beam, while Halls B and C wanted low current and polarized beam.
The wavelength of the laser chosen for the Hall~A system produced a high current beam 
with $\approx$~35\% polarization which is about half the beam polarization for Hall~C.
A 2\% leakage current from Hall~A dilutes the Hall~C beam polarization by about 1\%.
Since both Hall~B and C were at the same laser wavelength, the beam polarization at the injector
is the same for Halls~B and C. This means that Hall~B leakage current does not effect the
polarization of beam to Hall~C, but  changes to the relative amount from 
Halls~A and B to  the total leakage current in Hall~C does change the Hall~C beam polarization. There was
no measurement of the relative amount of leakage current from Halls~A and B in the
total measured leakage current. In addition to dilution from leakage current,
the longitudinal beam polarization at the Hall~C target depends on the energy per pass,
the number of passes and  the setting the spin rotator in the injector which was set to
maximize the product of longitudinal beam polarization in Halls~B and C. Therefore,
to accurately know the beam polarization in Hall~C, a measurement must be made near
the Hall~C target.

\begin{figure}[htb]
\includegraphics[width=\columnwidth]{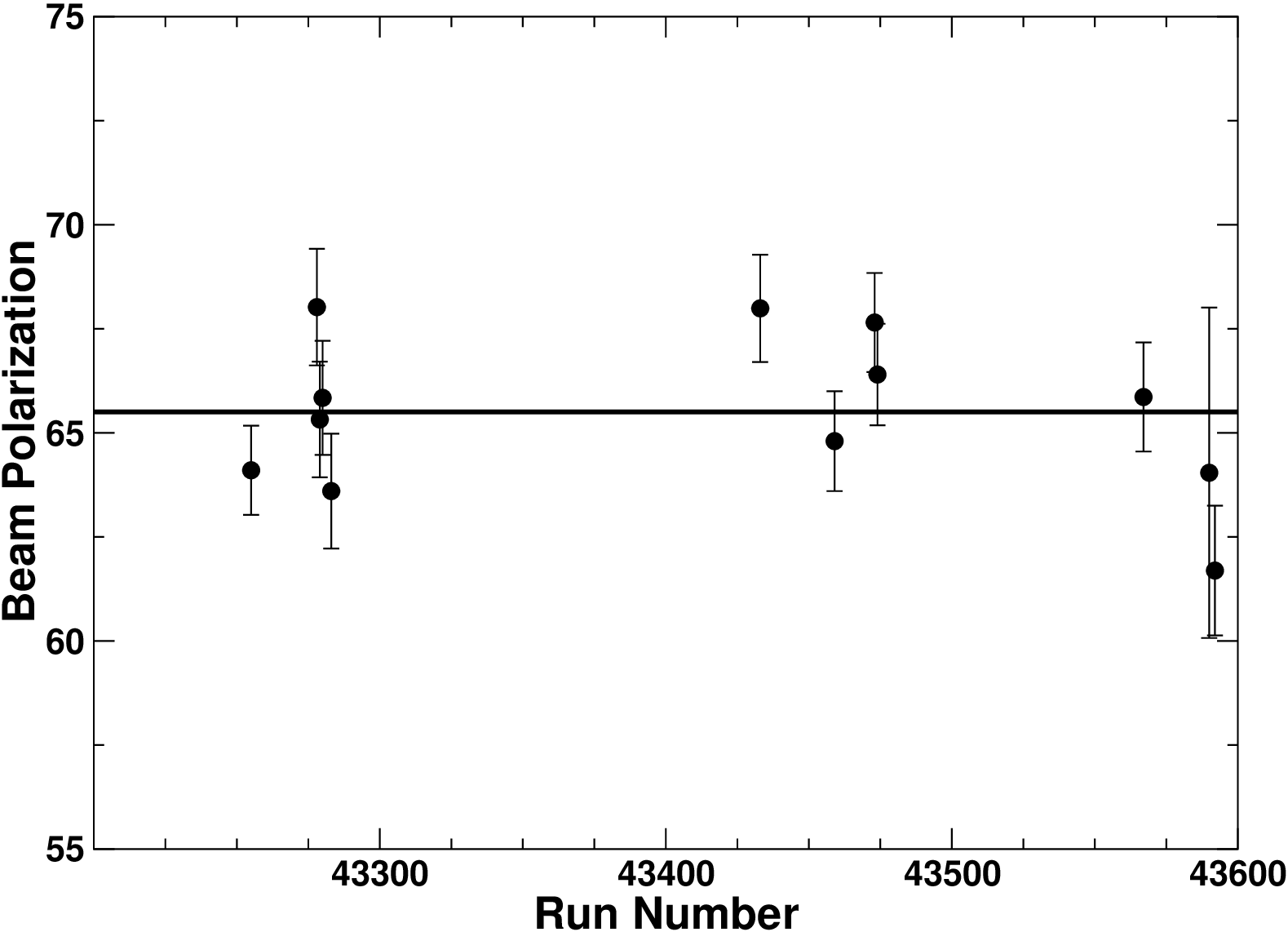}
\caption{
\label{fig:moller} 
Beam polarization, $P_B$, as a function of run number. The line is the 
weighted average of $P_B$.  
}
\end{figure}

The beam polarization, $P_B$,  was measured in Hall~C using the M\o ller
polarimeter~\cite{Ha99}. M\o ller measurements were taken when the target was retracted from the
beam for annealing. The M\o ller measurements were done at beam currents of 100 and 200~nA.
The measurements were taken throughout the run period and are plotted in Fig.~\ref{fig:moller}
as a function of run number. The average $P_B$ was $65.6 \pm 0.38$\% and was used to
determine the elastic asymmetry in Eq.~\ref{eq:asymexp}. The beam polarization was assumed to
be constant throughout the perpendicular target field running and no time dependent nor run-by-run
adjustment to the beam polarization was done.
The relative systematic error for the M\o ller measurement is 0.7\%. The beam polarization
could be different during the M\o ller measurements and the actual running due to changing
leakage currents in Halls~A and B. If the leakage current was mainly from Hall~B then
there would be no dependence of the Hall~C beam polarization on leakage current. The worse
case would be assuming that the leakage current is dominantly from Hall~A. With that condition
and assuming that leakage varies from 0\% to 4\% then an estimate of the relative systematic error on the 
beam polarization from changes in the leakage current is 1\%. Combining these 
errors in quadrature gives a relative systematic error of 1.3\% on the beam polarization.

When the target field is perpendicular to the beam direction, the incoming electrons
are bent downward before the target by the magnetic field. Two chicane magnets
before the target bend the incoming beam up so that, when combined with the target field, the
beam is incident horizontally on the target. The electrons scattered toward
 the HMS are bent downward and have an average out-of-plane angle of 3.4$^{\circ}$.

Normally, the position, angle and momentum of the scattered electron are determined
by measuring HMS focal plane position and angles of the electron and then reconstructing the
target quantities using an optics matrix. In addition, the HMS optics matrix
takes into account the vertical position of the beam at the target.
 The calculation of momentum and out-of-plane angle are sensitive to the vertical position. 
 The HMS optics matrix has been determined without  the target field.
The effect of the target field can be mimicked by using a effective vertical position at the target
with the known HMS optics matrix in an iterative procedure.
The reconstructed angles and momentum of the electron are determined 
using the known HMS optics matrix and an assumed effective vertical position at the target.
The electron is tracked from the entrance of the HMS back through the target field to 
the center of the target using  a tabulated map of the target field and
the reconstructed electron momentum and angle. The difference is taken between this 
tracked vertical position at the target center and the vertical position of the beam 
measured by the SEM. If the difference is larger than 1~mm, then a new effective vertical position
is assumed and the procedure is iterated until the difference between 
the tracked and measured vertical position is less than 1~mm.

To check the angle reconstruction, data were taken with the sieve collimator 
which has a 9x9 grid of holes. The pattern of sieve holes were properly
reconstructed by the algorithm described above. The momentum reconstruction was checked by
looking at the reconstructed final state mass, $W = \sqrt{M^2 + 2(E-E^{\prime})M - Q^2}$.
The peak position of $W$ was 
plotted as a function of different target variables. The $W$ peak position
had a slight dependence on the out-of-plane angle and no dependence on the other target variables. 
An azimuthal angle dependence was added to the map of the target field used in
the calculation of the electron's track which changed the electron's reconstructed momentum 
and eliminated the dependence of $W$ on the out-of-plane angle.

\section{Experimental Results}

From Ref.~\cite{Do86}, the beam-target asymmetry, $A_p$, for elastic electron-proton
scattering is related to the ratio of the proton's electric to magnetic
form factors, $r = G_E/G_M$,  by the formula:
\begin{equation}
A_p = \frac{-br\sin\theta^{\star}\cos\phi^{\star} - a\cos\theta^{\star}}{r^2+c} 
\label{eq:asym}
\end{equation}
in which $\theta^{\star}$ and $\phi^{\star}$ are the polar and azimuthal angles
between the momentum-transfer vector, $\vec{q}$, and the proton's spin vector.
$a,b,c$ are kinematic factors: 
\begin{eqnarray}
a &=& 2\tau\tan\frac{\theta_e}{2}\sqrt{1+\tau+(1+\tau)^2\tan^2\frac{\theta_e}{2}} \label{eq:a}\\
b &=& 2\tan\frac{\theta_e}{2}\sqrt{\tau(1+\tau)} \label{eq:b} \\
c &=& \tau + 2\tau(1+\tau)\tan^2\frac{\theta_e}{2} \label{eq:c}  
\end{eqnarray}
The measured asymmetry, $A_{m}$, is defined as  $(N^{+}-N^{-})/(N^{+}+N^{-})$ 
where $N^{+}$ and  $N^{-}$ are
the raw counts normalized  for deadtime and charge for opposite beam helicities.
The elastic asymmetry for the perpendicular target field is 
\begin{equation}
A_{p} = \frac{A_{m}}{f P_B P_T}  + N_{c} \label{eq:asymexp}
\end{equation}
where the measured asymmetry is normalized by $P_T$, $P_B$ and the 
dilution factor, $f$. The dilution factor is the ratio of 
the yield from scattering off free protons to that from the entire target.
$N_{c}$ is correction to 
the measured asymmetry which eliminates the contribution from 
quasi-elastic $^{15}$N scattering under the elastic peak.

\begin{figure}[h]
\includegraphics[width=\columnwidth]{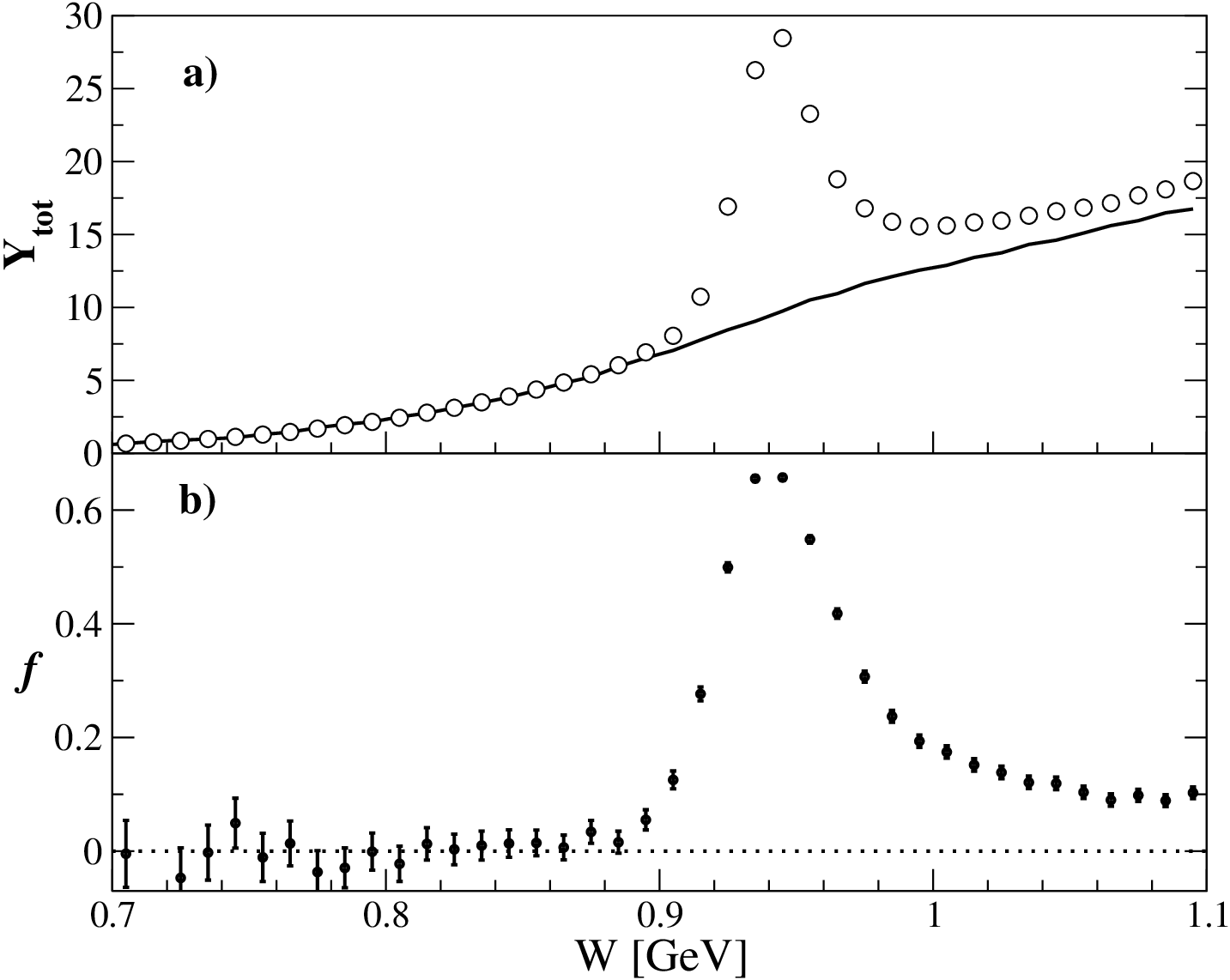}
\caption{
\label{fig:df} 
a) The yield, $Y_{tot}$, for scattering from the entire 
BOTTOM target is plotted as open circles versus W.
The error is smaller than the circle size. The solid line is $Y_{back}$, the $^{12}$C+He
yield which has been normalized to $Y_{tot}$ in the region of 0.6 $<$ $W$ $<$ 0.85~GeV. \\
b) The dilution factor, $f$, for the BOTTOM target versus W. The dotted line indicates zero
to guide the eye.
}
\end{figure}

 In Fig.~\ref{fig:df}a, the yield, $Y_{tot}$,
for scattering off the entire BOTTOM target is plotted versus W. The peak at $W$ $\approx$ 938~MeV
for elastic scattering off free protons is evident on top of the background from quasi-elastic
scattering from other target material. The width of the elastic peak is $\sigma = 14$~MeV and
is determined by the resolution in the scattered electron's momentum and angle. 
The width is consistent with a combination of  1.5~mr resolution in $\theta_e$
and  $1.5 \times 10^{-3}$ resolution in $E^{\prime}$. These resolutions are about 50\% larger than
the typical resolutions found with no target field and smaller raster size.  

To determine the shape of the quasi-elastic background
under the elastic peak, data were taken with a $^{12}$C 
disk (immersed in the liquid helium bath) of areal density
comparable to the ammonia in the target. The solid
line in  Fig.~\ref{fig:df}a is the yield, $Y_{back}$, from the  $^{12}$C+He data which has been normalized to
the BOTTOM target yield in the region 0.6 $<$ $W$ $<$ 0.85~GeV. The normalization factor was
$1.212 \pm 0.007$ for the BOTTOM target and $1.235 \pm 0.007$ TOP target. 
One can see that the $^{12}$C+He
matches the shape of $^{15}$N+He in the region  0.6 $<$ $W$ $<$ 0.85~GeV.  
The assumption that the shape of the
$^{12}$C+He is similar to the $^{15}$NH$_3$+He in the $W$ region under the elastic peak
was  tested by a Monte Carlo simulation using realistic cross section models and
including radiative corrections. The Monte Carlo predicts that normalization factor 
is 1.19 (1.22) for BOTTOM (TOP) target at $W$ = 0.77~GeV and has slight $W$ dependence 
of 0.04 every $\Delta W$ = 0.1~GeV. The difference in normalization factor
between the BOTTOM and TOP targets is caused by  different packing fractions (the ratio
of NH$_3$ to helium in the target).

The dilution factor, $f$, is $1-Y_{back}/Y_{tot}$ and $f$ for the BOTTOM target is plotted 
in  Fig.~\ref{fig:df}b. In the calculation of the dilution factor, the $W$ dependence 
of the normalization factor was not taken into account.
For W$ < 0.85$~GeV, $f$ is zero and flat indicating that the shape of the $^{12}$C+He data is well matched to
the shape of the $^{15}$N+He background with a constant normalization at all W$ < 0.85$~GeV. 
Near $W$ = 0.938~GeV, $f$ reaches a peak of about 0.66 and drops off to
near constant value of 0.10 for the $W$ region of the elastic radiative tail up to pion production threshold ($W = 1.075$~GeV). 
By combining the statistical error on the normalization factor and the
error due to assuming a flat $W$ dependence to the normalization factor, the relative systematic
error of 1.1\% on the dilution factor was calculated. 

\begin{figure}[h]
\includegraphics[width=\columnwidth]{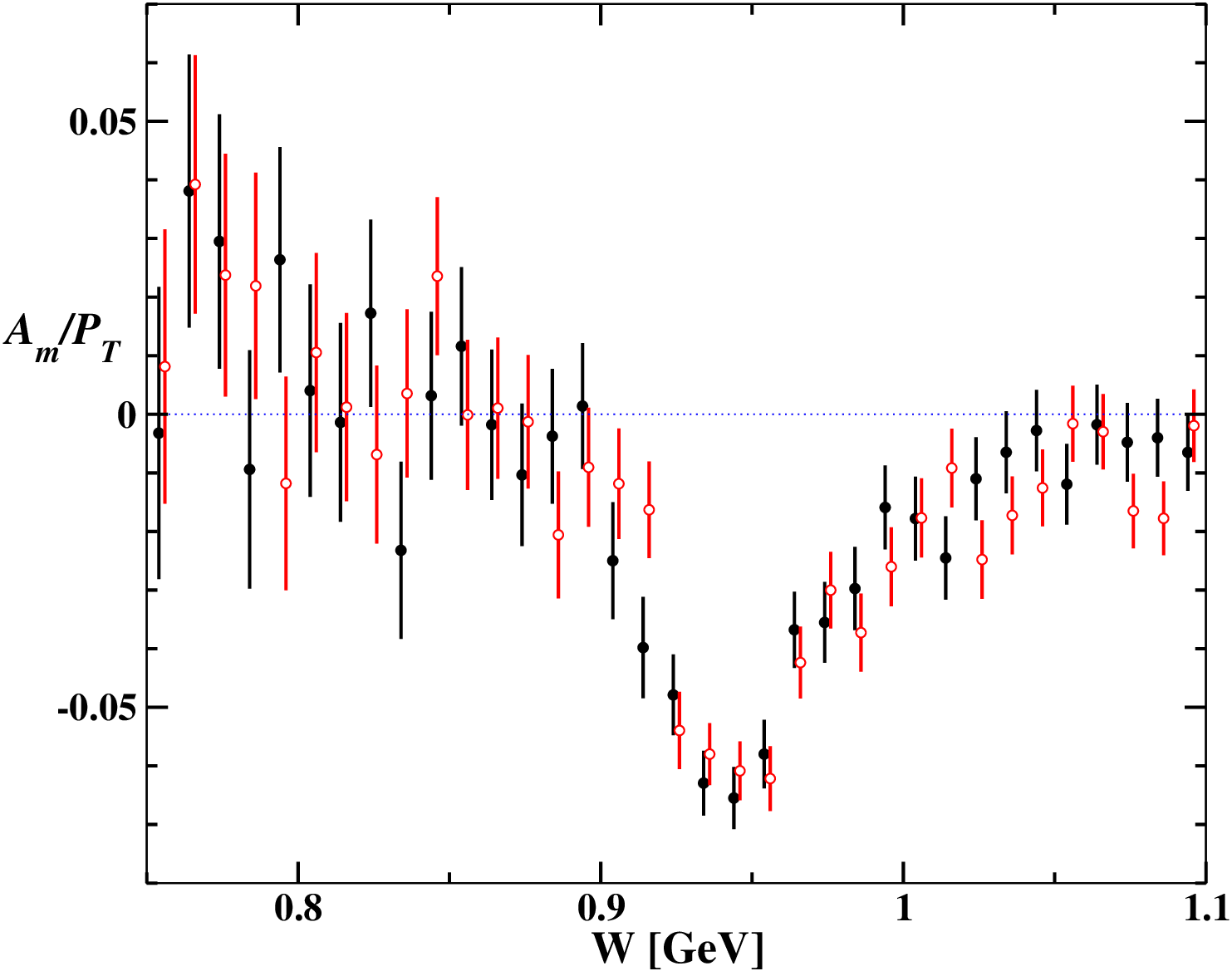}
\caption{ (Color online)
\label{fig:asymw} 
The asymmetry, $A_{m}/P^{ave}_T$, as a function of W. For the BOTTOM (TOP)
target, the asymmetry is plotted as a solid (open) circle. Each data set is
slightly shifted in $W$ for clarity.
}
\end{figure}

Typically, data taking was divided into runs of one hour duration
and $P_T$ changed during the run. $P_T$ was continuously measured and recorded
during the experiment every 20~seconds by an automated procedure.
The average proton polarization for all runs was 66\% (71\%) when running with the BOTTOM (TOP) target.
The charge-weighted average target polarization, $P^{ave}_{T}$, and  $A_{m}$ were measured for each run.
 In Fig.~\ref{fig:asymw}, the weighted average of 
$A_{m}/P^{ave}_T$ for all runs is plotted as a function of $W$ for BOTTOM
and TOP targets. 

The protons in $^{15}$N are polarized and contribute to $A_m$.
The contribution is characterized in terms of
the correction term, $N_c$, in Eq.~\ref{eq:asymexp}. $N_c$ 
 is equal to $f_{N}/f\times P_{N}/P{_T}\times {A_N}$
in which $f_N$, $P_N$ and $A_N$ are  the dilution factor, polarization and
asymmetry for the proton in  $^{15}$N. $A_N$ can be estimated from models~\cite{Ro99}.
From the angular momentum decomposition of the $p_{1/2}$ level that is populated
by the unpaired proton in the single particle shell model, one expects $A_N = -A_p/3$.
The polarization of the proton in $^{15}$N relative to $P_T$ 
has been measured in separate experiments~\cite{Cr95,Br90}. The data was fitted by the formula: 
\begin{equation}
P_N = -0.01\times(0.312+5.831|P_T|+8.935|P_T|^2 + 8.685|P_T|^3) \nonumber
\end{equation}
For $P_T = 71$\% one gets $P_N = -12$\%.
The dilution factor, $f_N$, is the ratio of the yield for scattering from
the polarized proton in $^{15}$N to the yield from scattering from the entire target. $f_N$ is 
like $f$ in that it varies with $W$ and  $f_N = 0.03$ at $W = 940$~MeV. The asymmetry
is corrected for $N_c$ at each $W$ and, to give a flavor of the size of the
correction, $N_c = -0.0002$ at $W = 940$~MeV which is a 0.2\% correction to $A_p$.

\begin{figure}[h]
\includegraphics[width=\columnwidth]{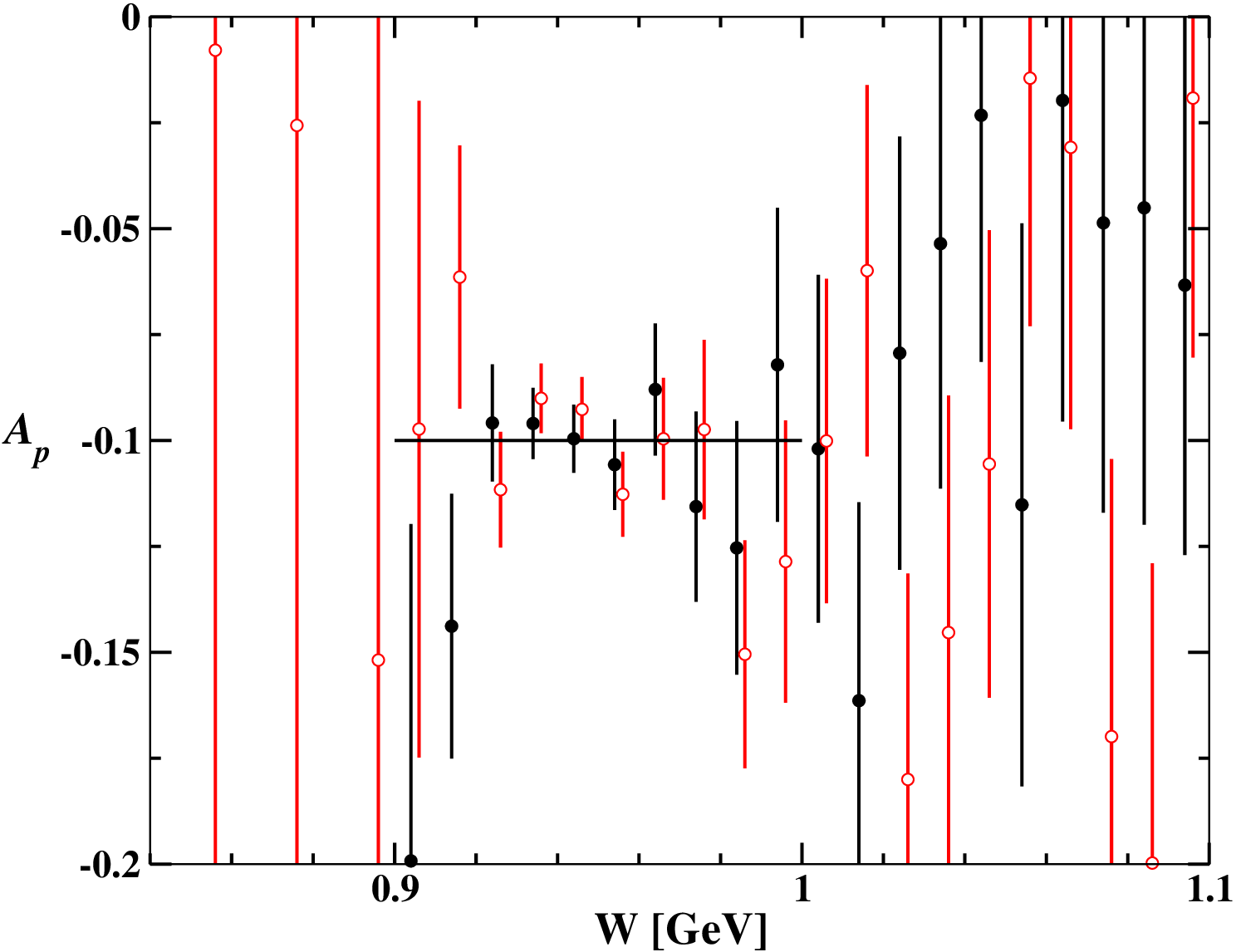}
\caption{ (Color online)
\label{fig:apw} 
The asymmetry, $A_{p}$, as a function of W. For the BOTTOM (TOP)
target, the asymmetry is plotted as a solid (open) circle. Each data set is
slightly shifted in $W$ for clarity. The solid line is the weighted average
of $A_p$ combining both targets.
}
\end{figure}

 $A_p$ is plotted  as a function of $W$ for both the BOTTOM
and TOP targets in Fig.~\ref{fig:apw}. For W$ < 0.9$~GeV,
$f$ is very small with relatively large error, so the error on $A_p$ becomes
larger than the scale of the y-axis. In the region $0.9 <$ $W$ $ < 1.0$~GeV, $A_p$ is
constant, and the error bars are small due to the large magnitudes of $A_m/P_t$ and $f$.  
 For W$ > 1.0$~GeV,
in the region of the elastic radiative tail, $A_{p}$ is still constant, but the error
bars are larger.
For the region $0.9 <$ $W$ $ < 1.0$~GeV, the average $A_{p}$
is $-0.1004 \pm 0.0042$ ($-0.0994 \pm 0.0044$) for BOTTOM (TOP) target. Radiative
corrections to $A_{p}$ were calculated using the MASCARAD code of Ref.~\cite{Af01} and  
shift $A_{p}$ by $-0.0004$. Including the radiative correction, the average $A_{p}$ from 
both targets is $-0.1003 \pm 0.0031$.

Using Eq.~\ref{eq:asym}, $G_E/G_M$ can be determined directly from $A_p$ using the formula:
\begin{eqnarray}
\frac{G_E}{G_M} = &-&\frac{b}{2A_p}\sin\theta^{\star}\cos\phi^{\star} \nonumber \\
&+&\sqrt{\frac{b^2}{4A^2_p}\sin^2\theta^{\star}\cos^2\phi^{\star}-\frac{a}{A_p}\cos\theta^{\star}-c}
\label{eq:gegm}
\end{eqnarray} 
in which $a,b$ and $c$ are the kinematic factors given in Eqns.~\ref{eq:a}-\ref{eq:c}. 
The average $\theta_e$ is 13.22$^{\circ}$ and the average $Q^2$ is 1.509~(GeV/c)$^2$.
The lab coordinate system is defined by the incoming and scattered electron's momentum vectors, 
$k$ and $k^{\prime}$, as positive $z$-direction along $\hat{k}$, $\hat{y} = \hat{k} \times \hat{k}^{\prime}$ 
and $\hat{x}  = \hat{y} \times \hat{z}$ with $+\phi$ rotation from  $+\hat{x}$ to  $+\hat{y}$. 
Since the scattered electron is bent downwards by the target's magnet field, 
the average azimuthal angle, $\phi_e$, is out-of-plane with a value of 348.8$^{\circ}$.
The $\vec{q}$ points  at the angles $\theta_q$ = 50.43$^{\circ}$ and $\phi_q$ = 168.8$^{\circ}$.
 For Eq.~\ref{eq:gegm},
one needs the polar and azimuthal angles, $\theta^{\star}$ and $\phi^{\star}$,
 between the $\vec{q}$ and the proton's spin vector.
Specifically, when the proton's spin vector is pointing at $\theta_s$ =90$^{\circ}$ 
and $\phi_s$ =180$^{\circ}$, $\theta^{\star}$ and $\phi^{\star}$ can be calculated by the
formulas:
\begin{eqnarray}
 \theta^{\star} &=& \arccos(\sin\theta_q\cos\phi_e) \nonumber \\ \nonumber
 \phi^{\star} &=& 180 + \arctan\left[\frac{\tan\phi_e}{-\cos\theta_q}\right]
\end{eqnarray}
For the present kinematics, $\theta^{\star}$ = 40.87$^{\circ}$ 
and $\phi^{\star}$ = 197.26$^{\circ}$. With these kinematic factors and the radiatively corrected 
average $A_p$, $\mu G_E/G_M$ $= 0.884 \pm 0.027$. The 
solution to Eq.~\ref{eq:asym} for $G_E/G_M$ is double-valued. 
The positive value of the square root was chosen, since the negative solution
gives an unreasonable value of $\mu G_E/G_M = -4.05$.
For this kinematic point, the systematic error on $\Delta (G_E/G_M)/(G_E/G_M)   = 0.97\times\Delta A_p/A_p$.
 The total relative systematic error on $\mu G_E/G_M$ is 3.3\%. A break down of the systematic
errors is given in Table~\ref{tab:sys}. The beam and target polarization
are the dominant contributions systematic contributions.

\begin{table}[bth]
\begin{ruledtabular}
\begin{tabular}{ccc}
Variable & Error & $\Delta r$/$r$ \\ \hline
$\theta_e$       & 0.5~mr  & 0.2\% \\
$\theta^{\star}$ & 0.1$^{\circ}$  & 0.1\% \\ 
$\phi^{\star}$   & 1.0$^{\circ}$   & 0.45\% \\ 
$E$              & 0.003~GeV & 0.005\%  \\ 
$E^{\prime}$     & 0.005~GeV & 0.01\%  \\ 
$f$              & 1.1\%       & 1.1\%  \\ 
$P_T$            & 2.9\%        & 2.8\%  \\ 
$P_B$            & 1.3\%        & 1.3\%  \\ \hline
Total            &             & 3.3\%  \\ 
\end{tabular}
\end{ruledtabular}
\caption[]{
Relative systematic errors on $r = G_E/G_M$. 
}
\label{tab:sys}
\end{table}

\section{Conclusion}

\begin{figure}[tbh]
\includegraphics[width=\columnwidth]{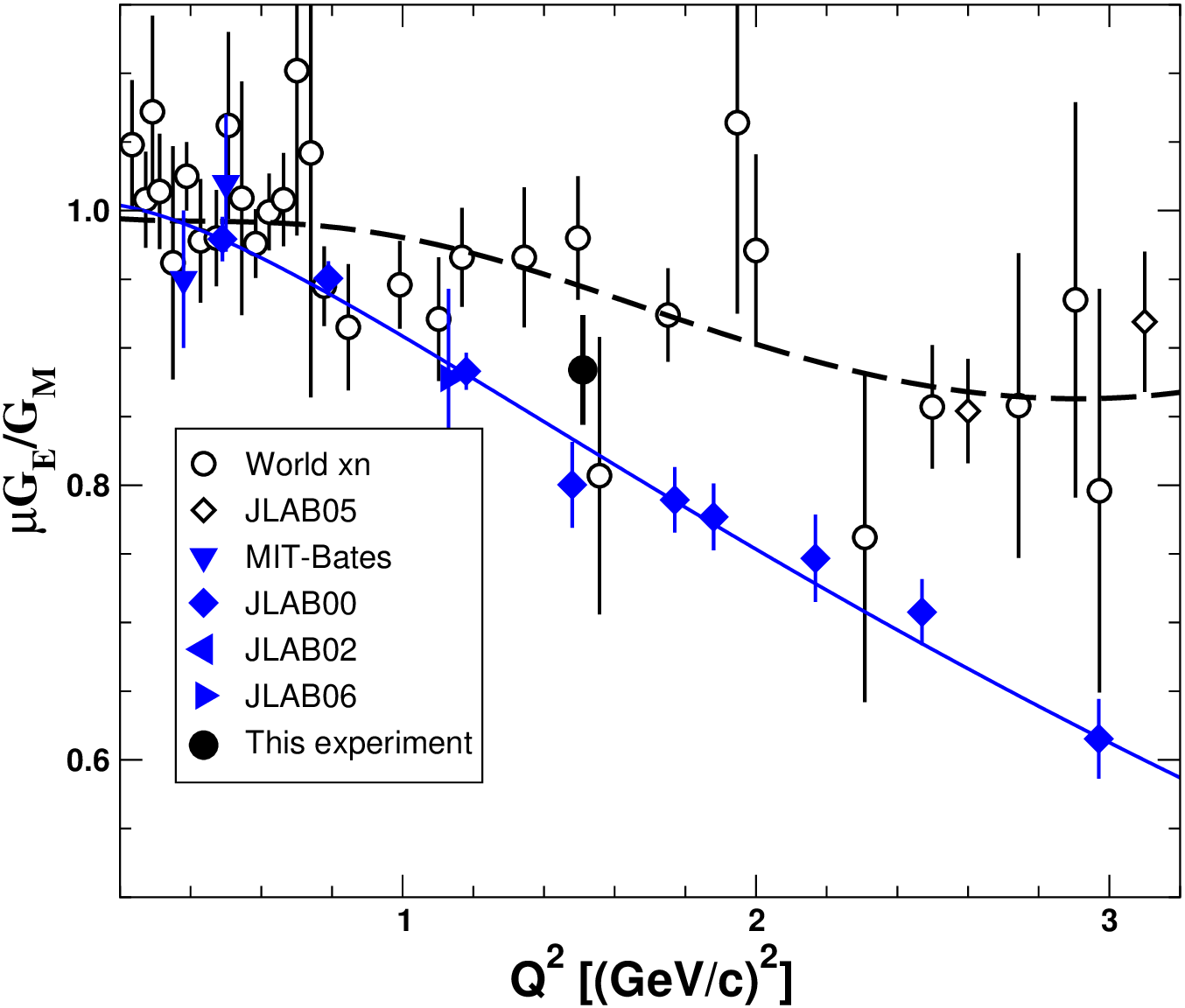}
\caption{
\label{fig:ratio} (Color online)
Ratio $\mu G_E/G_M$ plotted as a function of $Q^2$. The ratio $\mu G_E/G_M$ from this
experiment is plotted as a filled circle with the error bar being the
statistical and systematic error combined in quadrature. The solid line 
is a fit \cite{Lo02} to all form factor 
data, which only included proton $G_{E}/G_{M}$ from Ref.~\cite{Jo00} and \cite{Ga02}  
for large $Q^2$. Other symbols are same as in Fig.~\ref{fig:rat1}.
}
\end{figure}

In Fig.~\ref{fig:ratio},  the ratio $\mu G_E/G_M$ from this experiment is 
compared to previous measurements. 
A recent global fit of $G_E$ and $G_M$ to the world cross 
section data has been done~\cite{Ar04d}
and the result for $\mu G_E/G_M$  is plotted by a dashed line in Fig.~\ref{fig:ratio}.
The solid line is $\mu G_E/G_M$ from a fit to all nucleon form 
factors by Lomon~\cite{Lo02} which only uses proton $G_E/G_M$ from the polarization
transfer technique at large $Q^2$. The difference between the
two curves is 12\% at $Q^2$~=~1.509~(GeV/c)$^2$. The statistical error and 
systematic error for this measurement are comparable 
to previous $\mu G_E/G_M$ values from cross-section and recoil polarization
experiments. The data point is midway between the two curves
so it is about 2$\sigma$ away from either curve.
Unfortunately, the new measurement does not
help to determine whether the discrepancy between $\mu G_E/G_M$ from 
the Rosenbluth technique and the polarization transfer technique is due to unknown 
systematic errors in either technique.At this $Q^2$, inclusion the Coulomb distortion effects~\cite{Ar04c} 
in the Rosenbluth technique would reduce $\mu G_E/G_M$ by 0.05
which would make it overlap with the present data point and
bring measurements from all three techniques into reasonable agreement. 

The inclusion of two-photon exchange mechanisms in the extraction of the Born cross section
 will reduce $\mu G_E/G_M$ and bring it closer to
$\mu G_E/G_M$ determined by this measurement and previous measurements using the polarization transfer
technique. A calculation~\cite{Bl05} including all two-photon exchange mechanisms 
would reduce $\mu G_E/G_M$ by about 0.08
compared to the dashed line in Fig.~\ref{fig:ratio}. 
This beam-asymmetry measurement is 
at $\epsilon = 0.963$ which minimizes the contribution from two-photon exchange mechanisms
and, from Ref.~\cite{Bl05},  $\mu G_E/G_M$ would be reduced by roughly a factor of 0.995 by accounting
for the  two-photon amplitude mechanisms.

This experiment is the first to measure $G_E/G_M$ using  beam-target asymmetry 
in elastic $ep$ scattering.
To definitively distinguish between experimental techniques at this $Q^2$, a beam-target
asymmetry experiment  needs to reduce both the statistical and systematic error.   
The systematic error which is hardest to reduce is the error on the target polarization.
One approach would be to simultaneously measure the beam-target 
asymmetry at a given $Q^2$ with two separate spectrometers which are
at the same electron scattering angle but opposite sides of the beam.
By taking the ratio of the two asymmetry measurements, the beam and target polarization will cancel
and $G_E/G_M$ can be extracted with no systematic from either polarization
measurement.
Another approach would be to measure at
higher $Q^2$ where the percentage difference between $G_E/G_M$ extracted
from the two experimental techniques is larger, since the systematic error
on the beam and target polarization is independent of $Q^2$. To compensate for the falling
cross-section, the experiment either has to run longer or use large 
acceptance detectors to keep the statistical error from growing too large.
Dedicated experiments have been proposed \cite{Wa01,Zh04}
at Jefferson Lab to measure $G_E/G_M$ by beam-target asymmetries
using both these experimental approaches.

\begin{acknowledgments}
We would like to thank the Hall~C technical staff and
the accelerator operators for their efforts and dedication.
This work was supported by Schweizerische Nationalfonds, the
Department of Energy contract DE-FG02-96ER40950 and by the
Institute of Nuclear and Particle Physics of the University of Virginia.
The Southern Universities Research Association
(SURA) operates the Thomas Jefferson National Accelerator Facility for the
United States Department of Energy
under contract DE-AC05-84ER40150.

\end{acknowledgments}

\end{document}